\documentclass[twocolumn]{aastex61}

\usepackage[utf8]{inputenc} 
\usepackage[T1]{fontenc}    
\usepackage{hyperref}       
\usepackage{url}            
\usepackage{booktabs}       
\usepackage{amsfonts}       
\usepackage{nicefrac}       
\usepackage{microtype}      
\usepackage{lipsum}
\usepackage{graphicx}

\hypersetup{%
  colorlinks=true,
  linkcolor=blue,
  citecolor=blue,
  urlcolor=blue
}

\begin{document}
\shorttitle{Galactic continuum surveys} \shortauthors{Liu et al.}

\title{Galactic continuum surveys with the new 850 $\micron$ MKID camera at the EAO/JCMT 15-m telescope}
\correspondingauthor{Tie Liu}
\email{liutiepku@gmail.com}

\correspondingauthor{Di Li}
\email{dili@nao.cas.cn}

\author{Tie Liu}
\affiliation{East Asian Observatory, 660 N. A'ohoku Place, Hilo, HI 96720, USA}

\author{Di Li}
\affiliation{National Astronomical Observatories, Chinese Academy of Science, A20 Datun Road, Chaoyang District, Beijing 100012, China}
\affiliation{Key Laboratory for Radio Astronomy, Chinese Academy of Sciences, Nanjing 210008, China}

\author{David Eden}
\affiliation{Astrophysics Research Institute, Liverpool John Moores University, IC2, Liverpool Science Park, 146 Brownlow Hill, Liverpool L3 5RF, UK}

\author{James Di Francesco}
\affiliation{NRC Herzberg Astronomy and Astrophysics, 5071 West Saanich Rd., Victoria, BC V9E 2E7, Canada}
\affiliation{Department of Physics and Astronomy, University of Victoria, Victoria, BC V8P 1A1, Canada}

\author{JinHua He}
\affiliation{Key Laboratory for the Structure and Evolution of Celestial Objects, Yunnan Observatories, Chinese Academy of Sciences, P.O. Box 110, Kunming, 650011, Yunnan Province, People's Republic of China }
\affiliation{ Chinese Academy of Sciences, South America Center for Astrophysics (CASSACA), Camino El Observatorio 1515, Las Condes, Santiago, Chile}
\affiliation{Departamento de Astronomía, Universidad de Chile, Las Condes, Santiago, Chile}

\author{Sarah I. Sadavoy}
\affiliation{Harvard-Smithsonian Center for Astrophysics, 60 Garden Street, Cambridge, MA 02138, USA}

\author{Ken'ichi Tatematsu}
\affiliation{National Astronomical Observatory of Japan, 2-21-1 Osawa, Mitaka, Tokyo 181-8588, Japan}

\author{Kee-Tae Kim}
\affiliation{Korea Astronomy and Space Science Institute 776, Daedeokdae-ro, Yuseong-gu, Daejeon, 34055, Republic of Korea}

\author{Naomi Hirano}
\affiliation{Institute of Astronomy and Astrophysics, Academia Sinica. 11F of Astronomy-Mathematics Building, AS/NTU No.1, Sec. 4, Roosevelt Rd, Taipei 10617, Taiwan, R.O.C.}

\author{Gary Fuller}
\affiliation{UK ALMA Regional Centre Node, Jodrell Bank Centre for Astrophysics, School of Physics and Astronomy, The University of Manchester, Oxford Road, Manchester M13 9PL, UK}

\author{Yuefang Wu}
\affiliation{Department of Astronomy, Peking University, Beijing 100871, China}

\author{Sheng-Yuan Liu}
\affiliation{Academia Sinica, Institute of Astronomy and Astrophysics, P.O. Box 23-141, Taipei 106, Taiwan}

\author{Ke Wang}
\affiliation{The Kavli Institute for Astronomy and Astrophysics, Peking University, Beijing 100871, China}

\author{Mark Thompson}
\affiliation{Centre for Astrophysics Research, School of Physics Astronomy \& Mathematics, University of Hertfordshire, College Lane, Hatfield, AL10 9AB, UK}

\author{Mika Juvela}
\affiliation{Department of Physics, P.O.Box 64, FI-00014, University of Helsinki, Finland}

\author{Isabelle Ristorcelli}
\affiliation{IRAP, Universit\'{e} de Toulouse, CNRS, UPS, CNES, Toulouse, France}

\author{L. Viktor Toth}
\affiliation{E\"{o}tv\"{o}s Lor\'{a}nd University, Department of Astronomy, P\'{a}zm\'{a}ny P\'{e}ter s\'{e}t\'{a}ny 1/A, H-1117, Budapest, Hungary}

\author{Archana Soam}
\affiliation{SOFIA Science Centre, USRA, NASA Ames Research Centre, MS-12, N232, Moffett Field, CA 94035, USA}

\author{Patricio Sanhueza}
\affiliation{National Astronomical Observatory of Japan, National Institutes of Natural Sciences, 2-21-1 Osawa, Mitaka, Tokyo 181-8588, Japan}

\author{Sung-ju Kang}
\affiliation{Korea Astronomy and Space Science Institute 776, Daedeokdae-ro, Yuseong-gu, Daejeon, 34055, Republic of Korea}

\author{Woojin Kwon}
\affiliation{Korea Astronomy and Space Science Institute, 776 Daedeokdaero, Yuseong-gu, Daejeon 34055, Republic of Korea}
\affiliation{Korea University of Science and Technology, 217 Gajeong-ro, Yuseong-gu, Daejeon 34113, Republic of Korea}

\author{D. Ward-Thompson}
\affiliation{Jeremiah Horrocks Institute for Mathematics, Physics \& Astronomy, University of Central Lancashire, Preston PR1 2HE, UK}

 \author{Julien Montillaud}
\affiliation{Institut UTINAM - UMR 6213 - CNRS - Univ Bourgogne Franche Comte, France}

\author{John Richer}
\affiliation{Astrophysics Group, Cavendish Laboratory, J J Thomson Avenue, Cambridge CB3 0HE, UK}

\author{Jinjin Xie}
\affiliation{National Astronomical Observatories, Chinese Academy of Science, A20 Datun Road, Chaoyang District, Beijing 100012, China}

\author{Bingru Wang}
\affiliation{National Astronomical Observatories, Chinese Academy of Science, A20 Datun Road, Chaoyang District, Beijing 100012, China}

\author{Yapeng Zhang}
\affiliation{Department of Physics, The Chinese University of Hong Kong, Shatin, Hong Kong}



\begin{abstract}

This white paper gives a brief summary of Galactic continuum surveys with the next generation 850 $\mu$m camera at the James Clerk Maxwell Telescope (JCMT) in the next decade. This new camera will have mapping speeds at least 10-20 times faster than the present SCUBA-2 camera, and will enable deep ($<$10 mJy~beam$^{-1}$) and extensive continuum surveys such as a wider ($-5^{\circ}<l<240^{\circ}$, $|b|<2-5^{\circ}$) Galactic Plane survey, a larger Gould Belt survey, and a follow-up survey of all Planck compact objects visible from the northern hemisphere. These surveys will provide a complete census of molecular clouds, filaments, and dense cores across the Galaxy, vital for studying star formation in various environments.

\end{abstract}

\keywords{surveys---radio continuum: ISM---ISM: clouds---stars: formation}

\section{Introduction}
\label{sec:intro}

\renewcommand{\labelitemi}{\textasteriskcentered}

Stars form in molecular clouds when dense condensations of dust and gas gravitationally collapse to form a central protostar. This young stellar object (YSO) accretes material from the surrounding cloud material it reaches the temperature and density required for nuclear fusion. Knowledge of the physical factors that control the conversion of interstellar gas into molecular clouds, and then into stars, is of fundamental importance for understanding the star formation process and the evolution of galaxies. Nevertheless, our current theories of star formation are still very limited.

\begin{figure*}[tbh!]
  \centering
  \includegraphics[width=1\textwidth]{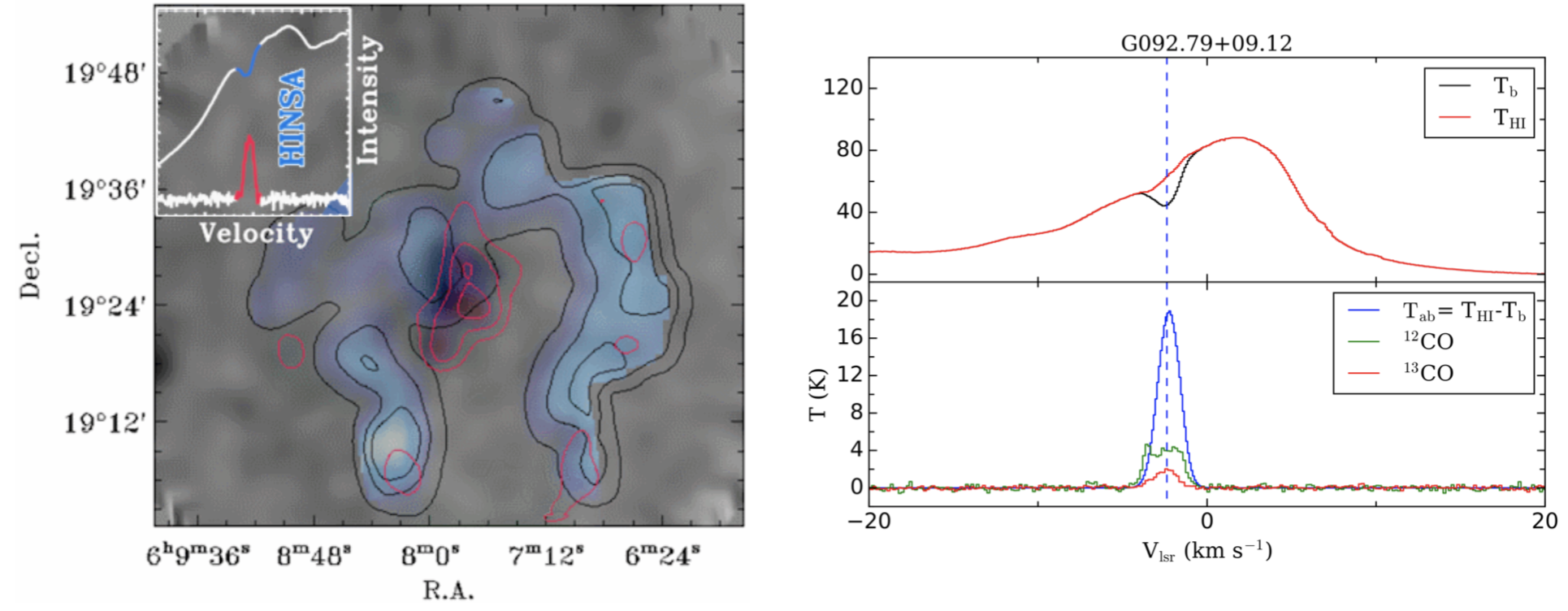}
  \caption{\textit{ \textbf{(Left:)} The B227 molecular cloud \citep{Zuo2018}. The gray scale image shows the 2MASS extinction; the black contours and blue shadow show the ratio of [HI]/[H$_2$]; and the red contours show the column density ratios of [$^{13}$CO]/[H$_2$]}.The upper left corner presents spectra of HI narrow self-absorption line and $^{13}$CO J=1-0 emission line.  \textbf{(Right:)} Spectra of PGCC G092.79+09.12 (Tang et al. 2019 submitted). Top panel: black and red solid lines are the observed HI and the recovered HI background spectrum $T_{\rm HI}$, respectively. Bottom panel: the profile of absorption temperature $T_{ab}$ of HI (blue), $^{12}$CO J=1-0 (green) and $^{13}$CO J=1-0 spectrum (red).}
  \label{fig:HI}
\end{figure*}

Molecular clouds form stars in their deepest, high-extinction interiors. Such regions only account for small percentages of the sizes and masses of molecular material in these clouds. The key to understanding the formation of clouds themselves may lie in sensitive observations of their diffuse outermost extents. The outermost extents of clouds, where HI is converted to H$_2$ (and vice-versa) are difficult to probe. Through a combined analysis of HI narrow self-absorption, CO emission, dust emission, and extinction, \cite{Zuo2018} identified a striking "ring" of enhanced HI abundance in a molecular cloud (see left panel in \textit{Figure} \ref{fig:HI}), resembling the
``onion'' shell description of a forming molecular cloud with ongoing H$_2$ formation. Such observations, however, are very rare, preventing us from a thorough understanding of cloud formation and evolution.

Recent studies have defined a ``threshold'' surface density of about 120 M$_{\odot}$~pc$^{-2}$ for star formation in nearby clouds \citep{Heiderman2010,Lada2010,Lada2012}, above which the vast majority of dense cores and YSOs are found. \cite{Evans2014} compared various models of star formation to observations of nearby clouds and found that the mass of dense gas was the best predictor of the star formation rate (SFR). More recently, \cite{Vutisalchavakul2016} found a similar threshold applied to more distant and massive clouds in the Galactic Plane, using millimeter-continuum emission from the Bolocam Galactic Plane Survey (BGPS) \citep{Ginsburg2013} to measure the mass of dense gas. They further demonstrated that the average star formation rate per unit mass of dense gas (star formation efficiency; SFE) is remarkably constant over a large range of scales and conditions, ranging from Galactic molecular clouds to clouds in nearby galaxies to more distant galaxies \citep{Vutisalchavakul2016}. 

However, the particular threshold or constant SFE may not apply in some extreme regions, especially those with lower metallicity or much stronger radiation fields (e.g. clouds in the central molecular zone,CMZ, \citep{Longmore2013}). In particular, the fraction of dense gas in various kinds of clouds may highly depend on cloud properties. Whether or not there is a single, universal density threshold for star formation can only be tested by
investigating a complete and unbiased volume-limited sample of star-formation activity not only in the well-known dark clouds of the Galactic plane and Gould's Belt but also those currently unknown regions, potentially located in diffuse or small molecular clouds.

Whilst there is a broad-brush theory behind (low-mass) star formation, the initial conditions that lead a molecular cloud core to collapse and ultimately form a star is still far from fully understood. This limitation has a serious impact on our capability to extend the models of current star formation in our Galaxy to other environments such as low-metallicity galaxies or starburst galaxies. In the current paradigm, stars form within the compact (with sizes of 0.1~pc or less), cold ($T_{\rm k}\leq$10 K), and dense ($n$(H$_{2}$)$>5\times10^{4}$ cm$^{-3}$) starless fragments in molecular clouds, usually dubbed ``prestellar cores" \citep{Caselli2011}. The prestellar core mass function in nearby clouds is very similar in shape to the stellar initial mass function \citep{Andre2014,Konyves2015}, hinting that there is a direct physical link between the initial conditions in prestellar cores and the star formation process (e.g. star formation mode (clustered or isolated), star formation efficiency and rate). However, it is unclear whether or not the core mass function is universal in clouds or if it changes across different environments.

Recent studies of nearby clouds by Herschel revealed a ``universal" filamentary structure in the cold ISM \citep{Andre2014}. Filaments are commonly surrounded by a network of perpendicular striations. 
Filaments have been well predicted in simulations of supersonic turbulence with the absence of gravity, which can produce hierarchical structure with a lognormal density distribution seen in observations \citep{Vazquez-Semadeni1994}. Filaments in strongly magnetized turbulent clouds (e.g. B211/3, \citep{Palmeirim2013}) are formed preferentially perpendicular to the magnetic field lines, suggesting that magnetic fields have an important role in filament formation. \cite{Li2019} showed that a moderately strong magnetic field ($\mathcal{M_A}\sim1$) is crucial for maintaining a long and slender filamentary cloud for a long period of time $\sim$0.5 million years.  Polarization observations of nearby filamentary clouds suggested a scenario in which local interstellar material has condensed into a gravitationally-unstable filament (with “supercritical” mass per unit length) that is accreting background material along field lines through the low-density striations \citep[e.g.,][]{Cox2016}. The formation of filaments by self-gravitational fragmentation of sheet-like clouds was also seen in simulations of 1D compression (e.g. by an expanding H{\sc ii} region, an old supernova remnant, or collision of two clouds). For example, the asymmetric column density profiles of filaments in the Pipe Nebula have been attributed to the large-scale compressive flows generated by the winds of the nearby Sco OB2 association \citep{Peretto2012}. Cloud-cloud collisions may also trigger the formation of these dense filaments \citep{Wu2017}. Although significant progress has been made in recent years, how filaments form in the cold ISM is still poorly understood.  

The properties of filaments themselves are also far from well understood. One of the most striking results from Herschel observations in the Gould Belt clouds is the finding of an apparent characteristic width ($\sim$0.1 pc) of filamentary substructures \citep{Andre2014}. The origin of such a characteristic width is not well understood  and this result is highly debated. Projection effects or artifacts in the data analysis may affect the inferred filament width \citep{Panopoulou2017}. Some numerical simulations modeling the interplay between turbulence, strong magnetic field, and gravitationally driven ambipolar diffusion are indeed able to reproduce filamentary structures with widths peaked at 0.1 pc over several orders of magnitude in column density \citep[e.g.,][]{Auddy2016,Federrath2016}.  Other simulations, however, find a wide range of filament widths \citep[e.g.][]{Smith2014}. It is unclear whether or not filaments have a universal width or under what conditions the width of a filament would deviate from 0.1 pc. Since we expect turbulence, magnetic fields, and external pressure to shape clouds and their filaments, we need a large census of filament properties and their host cloud properties to investigate how these filaments form and how they will in turn influence the types of cores and stars that will be produced. \\

\section{Current Status}
\label{sec:current}

Prestellar cores are located in the coldest parts of molecular clouds. Early infrared surveys (e.g. IRAS, MSX, Spitzer, WISE) traced warm
dust emission associated with protostellar objects or more evolved YSOs that are too old to assess the initial conditions of star formation.
Extinction studies have assessed cloud structure and mass distributions, but these studies are mostly limited to low-latitude clouds where there are sufficient background stars to assess the dust content in the foreground clouds.
Due to optical depth and depletion effects, molecular gas surveys (e.g., CO) are poor tracers of the densest parts of molecular clouds. Instead, the best probe of molecular cloud structures from the diffuse cloud material to the small prestellar cores and the entire star formation process is from thermal dust emission of cold dust grains. We briefly summarize several representative (sub)mm continuum surveys in \textit{Table \ref{Tab}}.  

\begin{table*}[tbh]
\setlength{\abovecaptionskip}{0pt}
\setlength{\tabcolsep}{0.02in}
\begin{center}
\caption{\textbf{List of representative (sub)millimeter continuum surveys in the Galaxy}}
\begin{tabular}{cccccc}
\hline\hline \noalign {\smallskip}
Survey & Bands & Angular resolution  & Survey region ($^{\circ}$) & Sky Coverage & rms (mJy/beam)\\
\hline \noalign {\smallskip}
BGPS/CSO & 1.1 mm & 33$^{\prime\prime}$ &   $-10.5<l<90.5, |b|<0.5-1.5$ & 170 deg$^{2}$ & 11-53\\
ATLASGAL/APEX & 870 $\mu$m & 17$^{\prime\prime}$.5  &  $-80<l<60, |b|<1-2$ & $\sim$400 deg$^{2}$ & 40-50\\
Hi-GAL/Herschel  & 70-500 $\mu$m & 5$^{\prime\prime}$-35$^{\prime\prime}$ & $-180<l<180, |b|<1$ & $\sim$720 deg$^{2}$ \\
GBS/Herschel  & 70-500 $\mu$m & 5$^{\prime\prime}$-35$^{\prime\prime}$ & Gould Belt clouds & $\sim$160 deg$^{2}$ \\
SASSy/JCMT  & 850 $\mu$m & 14$^{\prime\prime}$.1  &  $60<l<240, |b|<2$  & $\sim$720 deg$^{2}$ & 25-40\\
JPS/JCMT  & 850 $\mu$m & 14$^{\prime\prime}$.4  &  $7<l<63, |b|<1$  & $\sim$50 deg$^{2}$ & 8.4 \\
GBS/JCMT  & 450-850 $\mu$m & 7$^{\prime\prime}$.5-14$^{\prime\prime}$.5  &  Gould Belt clouds  & $\sim$24 deg$^{2}$ & 5 \\
SCOPE  & 850 $\mu$m & 14$^{\prime\prime}$.1  & 1235 Planck Galactic Cold Clumps & $\sim$50 deg$^{2}$ & 6-10\\
Planck    & 0.35-10 mm    & $\sim$5$^{\prime}$          & All sky             & $\sim$41253 deg$^{2}$ \\
\hline \noalign {\smallskip}
\end{tabular}
\label{Tab}
\end{center}
\end{table*}

\subsection{Galactic Plane surveys}

The Herschel infrared Galactic Plane Survey \citep[Hi-GAL,][]{Molinari2010} observed the entire plane in five continuum bands between 70 $\mu$m and 500 $\mu$m. This survey provides the far-infrared and submillimeter continuum data to conduct self-consistent measurements of dust temperature and dust mass throughout the entire Galactic Plane, from the CMZ to the outer most regions of the Galaxy. Hi-GAL DR1 catalogues of the inner Milky Way contain 123210, 308509, 280685, 160972, and 85460 compact sources in the five bands \citep{Molinari2016}. Elia et al. (2017) studied the physical parameters of those compact sources with respect to Galactic longitude and their association with spiral arms. They found only minor or no differences between the average evolutionary status of sources in the fourth and first Galactic quadrants, or between `on-arm' and `interarm' positions \citep{Elia2017}. They also found that the surface density of sources increases as they evolve from prestellar to protostellar, but decreases again in the majority of the most evolved clumps.

The Bolocam Galactic Plane Survey (BGPS) is a 1.1 mm continuum survey at 33" effective resolution of 170 deg$^{2}$ of the Galactic Plane visible from the northern hemisphere \citep{Aguirre2011}. It was conducted with the Bolocam instrument on the 10.4 m Caltech Submillimeter Observatory. The survey has detected approximately 8600 clumps over the entire area to a limiting non-uniform 1$\sigma$ noise level in the range 11-53 mJy~beam$^{-1}$ in the inner Galaxy \citep{Rosolowsky2010,Ginsburg2013}. The majority of BGPS sources are molecular clumps.  These clumps are large dense subregions in molecular clouds with supersonic turbulence \citep{Shirley2013} that are linked to cluster formation \citep{Rosolowsky2010}. \cite{Svoboda2016} identified a subsample of 2223 (47.5\%) starless clump candidates (SCCs) from BGPS objects and found that $>$75\% of SCCs with known distances appear gravitationally bound.

The APEX Telescope Large Area Survey of the Galaxy (ATLASGAL) is one of the largest and most sensitive (rms$\sim$40 to 50 mJy~beam$^{-1}$) ground-based submillimetre (870 $\mu$m) surveys of the inner Galactic Plane \citep{Schuller2009}. The ATLASGAL Compact Source Catalogue \citep[CSC;][]{Contreras2013,Urquhart2014} includes $\sim$10,000 dense clumps. Urquhart et al. (2018) found that the vast majority of the detected clumps are capable of forming high-mass stars and 88 per cent are already associated with star formation at some level \citep{Urquhart2018}. They also found that all of the clumps are tightly correlated with the mid-plane of the Galaxy with a scale height of $\sim$26 pc. Li et al. (2016) found that almost 70\% of the total 870 $\mu$m flux associated with the coherent structures in ATLASGAL survey resides in filaments \citep{Li2016}. They also found that these filamentary structures are tightly correlated with the spiral arms in longitude and velocity, and that their semi-major axis is preferentially aligned parallel to the Galactic mid-plane, and therefore with the direction of large-scale Galactic magnetic field \citep{Li2016}.

The James Clerk Maxwell Telescope (JCMT) Plane Survey \citep[JPS;][]{Eden2017,Moore2015} was originally planned to observe the entire Galactic Plane, but was scaled back to be a targeted, yet unbiased survey of the Inner Galactic Plane. The survey strategy consisted of observing six equally spaced fields of approximately 5$^{\circ}$ in longitude, 1$^{\circ}$.7 in latitude centred at longitudes of l = 10$^{\circ}$, 20$^{\circ}$, 30$^{\circ}$, 40$^{\circ}$, 50$^{\circ}$, and 60$^{\circ}$ and a latitude of b=0$^{\circ}$. The rms obtained was at least 8.42 mJy~beam$^{-1}$.  A compact-source catalogue was produced for the JPS \citep{Eden2017} resulting in 7813 sources, with 95 per cent completeness limits estimated to be 40 mJy~beam$^{-1}$ and 300 mJy for the peak and integrated flux densities, respectively. The increased resolution of the JPS survey breaks up sources in the ATLASGAL survey, allowing for a more accurate identification of cold, dense, star-forming clumps. Using the JPS 850 $\mu$m-continuum images in a larger-scale study, maps of the dense-gas mass fraction were produced, and a power-spectrum analysis found increases on the scales of individual molecular clouds (Eden et al., in preparation). Looking at individual regions, the JPS data finds that W43 appears to be a younger star-formation source than the rest of the Milky Way \citep{Moore2015}. The star-forming regions of W49 and W51 were found to have consistent clump-mass functions but the luminosity function of W49 was found to be flatter, indicating it may be a candidate starburst region \citep{Eden2018}. Star-forming clumps were also found to be more centrally concentrated than those not housing a YSO, indicating that star formation is altering the morphology of a clump \citep{Eden2017}.

The SCUBA-2 Ambitious Sky Survey \citep[SASSy;Thompson et al., in preparation;][]{MacKenzie2011,Nettke2017} at the JCMT was originally designed as an all-sky survey with a goal of mapping the entire sky visible from the JCMT at a consistent rms. However, as with the JPS, this goal was scaled back and resulted in two surveys of the Outer Galaxy, SASSy-Perseus and SASSy-Outer-Galaxy with rms values of 25-40 mJy~beam$^{-1}$. The combined surveys contain 3138 sources and provides a nice complement to the ATLASGAL survey that has similar sensitivity limits in the longitude range 300$^{\circ} < l < 60^{\circ}$ \citep{Contreras2013,Urquhart2014}. A recent SASSy result extended the above results to greater Galactocentric radii, looking at the star-formation efficiency in the Outer Galaxy.  They find that the star-formation efficiency decreases with distance from the CMZ \citep{Djordjevic2019},  in contrast to previous studies looking at nearby clouds that found a constant star formation efficiency. 

\subsection{Gould Belt Clouds}

Herschel/GBS is an extensive imaging survey of the densest portions of the Gould Belt with SPIRE at 250-500\,$\mu$m and PACS at 110-170\,$\mu$m. The main findings of Herschel/GBS are: (1) Herschel reveals a ``universal'' filamentary structure in the cold ISM \citep[and references therein]{Andre2014}; (2) more than 70\% of the prestellar cores and embedded protostars identified with Herschel are found within the densest filaments, i.e., those with column densities exceeding $\sim7\times10^{21}$ cm$^{-2}$ \citep[and references therein]{Andre2014}; (3) the prestellar core mass function is very similar in shape to the stellar initial mass function \citep{Andre2014,Konyves2015}.

The JCMT Gould Belt Survey (GBS) was one of the first large legacy programs approved at the JCMT.  With the aim of characterizing nearby star formation, the GBS utilized hundreds of hours to map thermal dust emission at 850 $\mu$m, as well as line emission over a more focused area, within nearby molecular clouds up to 500 pc in distance. Science topics addressed by the survey include the evolution of dust grains \citep[e.g.,][]{Sadavoy2013}, the virial properties of dense cores \citep[e.g.,][]{Pattle2015}, the influence of protostellar heating on dense cores \citep[e.g.,][]{Rumble2015}, the clustering properties of dense cores \citep[e.g.,][]{Kirk2016}, and the properties of filamentary structure \citep[e.g.,][]{Salji2015}.  Work on a final GBS catalogue of star-forming cores in Gould Belt clouds is ongoing and will be published in 2020.  The GBS data are also providing an important dataset for a wide variety of other large surveys including the BISTRO polarization survey \citep{Ward-Thompson2017}, the Green Bank Ammonia Survey \citep{Friesen2017}, and the JCMT Transient Survey \citep{Herczeg2017}, as well as individual PI-led surveys such as an ALMA search for substructure within dense cores in the Ophiuchus molecular cloud \citep{Kirk2017}.

\subsection{The SCOPE survey of Planck Galactic Cold Clumps}

\begin{figure*}[tbh!]
  \centering
  \includegraphics[width=1\textwidth]{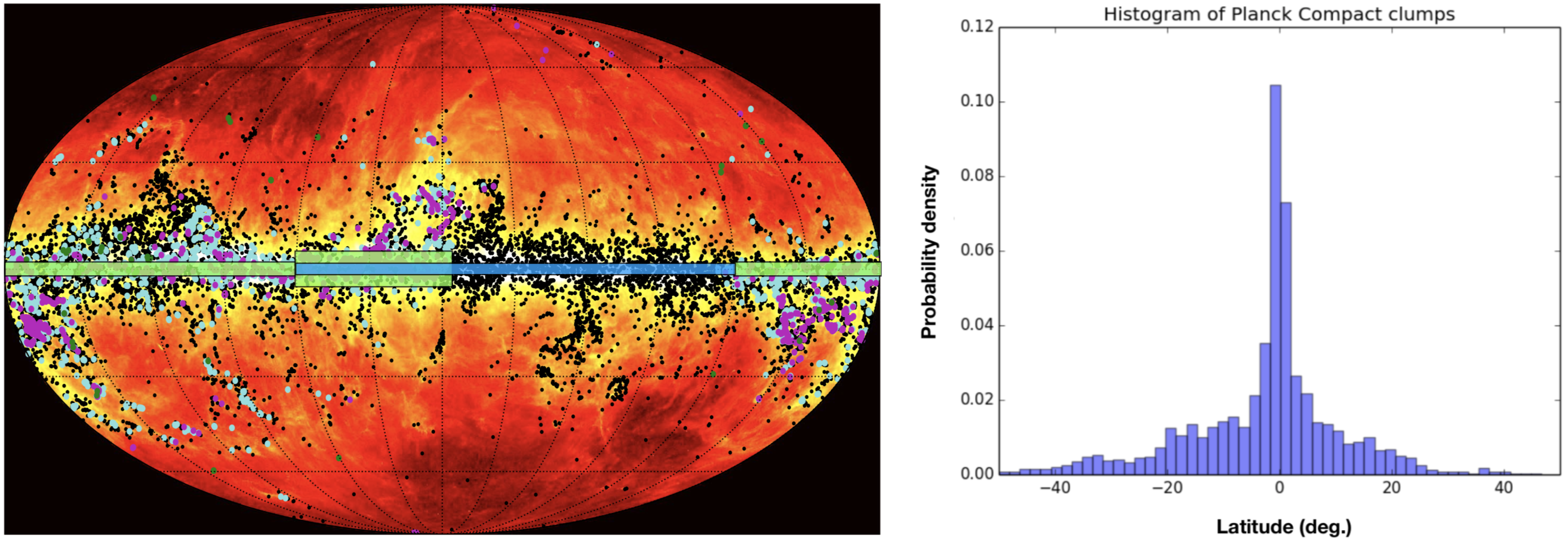}
  \caption{\textit{ \textbf{(Left:)} All-sky distribution of the 13188 PGCC sources (black dots), the 2000 selected PGCC sources in the TOP survey (blue dots) and the 1200 PGCCs selected in the SCOPE survey overlaid on the 857 GHz Planck map. The green box outlines the sky coverage of the proposed JCMT Galactic Plane survey with the new 850 $\mu$m camera. The blue box outline the sky coverage of the ATLASGAL survey. \textbf{(Right:)} Distribution of Planck compact sources at 353 GHz as a function of Galactic latitude.} }
  \label{fig:SCOPE}
\end{figure*}

Planck is the third generation mission to measure the anisotropy of the cosmic microwave background (CMB), and it observed the sky in nine frequency bands (the 30, 44, 70, 100, 143, 217, 353, 545, and 857 GHz bands). The high frequency channels of Planck cover the peak thermal emission frequencies of dust colder than 14 K \citep{Planck2011,Planck2016}, indicating that Planck could probe the coldest parts of the ISM. There are 13188 cataloged Planck galactic cold clumps (PGCCs), which spread across the whole sky, i.e., from the Galactic plane to high latitudes, following the spatial distribution of the main molecular cloud complexes. The low dust temperature ($<$14 K) of PGCCs makes them likely to be pre-stellar objects or at the earliest stages of protostellar collapse \citep{Planck2011,Planck2016}. The SCUBA-2 Continuum Observations of Pre-protostellar Evolution survey (SCOPE) is a legacy survey using SCUBA-2 at the JCMT to survey $\sim$1000 PGCCs at 850 $\mu$m at higher resolution than what can be obtained with Planck, aiming to conduct of census of cold dense cores in widely different environments \citep[see left panel of \textit{Fig. \ref{fig:SCOPE}};][]{Liu2018a,Eden2019}.  SCOPE has resolved the PGCCs into more than 3000 dense cores, many of which are located in high-latitude ($|b|>10^{\circ}$) or low density ($N<5\times10^{21}$ cm$^{-2}$) clouds \citep{Eden2019}. Follow-up observations of the dense cores discovered in the SCOPE survey suggest that most of them are either starless cores or very young Class 0/I objects \citep{Tatematsu2017,Liu2016,Liu2018b,Yi2018}, representing the earliest phases in pre-/proto-stellar evolution.

\subsection{The need of new Galactic (sub)mm continuum surveys}

The aforementioned previous surveys were limited by either low resolution (like Planck) or poor sensitivity (like BGPS, ATLASGAL, SASSy). Except for the all-sky Planck survey, all other previous continuum surveys have small sky coverage ($<$2\% of the sky) and do not include high latitude or mid-latitude clouds. As shown in Table \ref{Tab}, most of the previous (sub)mm continuum surveys focused on Gould Belt clouds or the inner Galactic Plane with either limited coverage or limited resolution. A major portion of the Galaxy (e.g. intermediate-latitude or high-latitude clouds) has not been fully explored at (sub)millimeter bands with appropriate sensitivity and high angular resolution ($\times$10$^{\prime\prime}$), which limits our knowledge of star formation and cloud evolution across the whole Galaxy. 

These well studied Gould Belt clouds do not broadly represent the star formation activity seen in regions like metal poor high-latitude clouds, the CMZ of the Milky Way, or starburst galaxies, where densities, temperatures, and star formation efficiencies can be vastly different. There are still fundamental aspects of the initial conditions for star formation that remain unaddressed, which include but are not limited to:

$\bullet$ On small scales, how and where do prestellar cores (i.e. future star forming sites) form? Specifically, can prestellar cores form in less dense and metal poor high-latitude clouds, or short lived cloudlets? What is the interplay between turbulence, magnetic field, gravity, kinematics and external pressure in prestellar core formation and evolution? Is there really a ``universal'' density threshold or core mass function for core/star formation?

$\bullet$ On larger scales, how common are filaments in molecular clouds? How do they form? How do they fragment and what is their role in producing prestellar cores?  Do filaments have a universal width?

$\bullet$ On cloud scales or galaxy scales, how do molecular clouds form from the diffuse atomic gas? Is there a universal dense-gas star formation law for clouds in various environments (e.g. spiral arms, interarm regions, high latitude, expanding H{\sc ii} regions, supernova remnants)? What factors determine and regulate star formation rates and efficiencies?\\

To address these questions, a census of a broader representation of star formation in the Galaxy is needed. That broader representation requires going beyond the dense gas in nearby clouds or toward the Galactic Plane to the entire spectrum of molecular clouds throughout the Galaxy. We want to compare and contrast star formation in clouds located in the CMZ, in spiral arms, in interarm regions, at high latitudes, and at high Galactocentric radii to span a range of ISRFs, metallicities, pressures, and ionizations.   

We have realized that the present SCUBA-2 instrument lacks the sensitivity and speed to conduct such large, unbiased surveys of all of these cloud regimes. Therefore, a new camera with much improved sensitivity and faster mapping speed is highly demanded. 

\section{The Next Decade}
\label{sec:future}

\subsection{The new 850 $\mu$m MKID camera}

The detector technology for the next generation 850 $\mu$m wide field camera is Microwave Kinetic Inductance Detectors (MKID), with intrinsic polarization mapping capability. These devices are much simpler to manufacture and operate than the TES detectors used in SCUBA-2 and are already being deployed on other large mm/submm telescopes such as the IRAM 30-m Telescope. The full field of view (FoV) of the new 7,272 MKID array will be 12 arcmin in diameter (almost twice as large as SCUBA-2). The focal plane will be filled with 3,636 1 f$\lambda$ spaced feedhorn coupled pixels. More details of the new instrument can be found on the JCMT webpage\footnote{https://www.eaobservatory.org//jcmt/wp-content/uploads/sites/2/2019/05/Guide-to-the-new-850um-MKID-camera-performance.pdf}

\subsubsection{The capability of the new camera: Much faster scan speeds}

The per-pixel sensitivity on the sky of the new camera will be a factor of 3 more sensitive than SCUBA-2. Between a larger FoV that almost doubles the exposure area in a larger Pong map and the more sensitive pixels, the new instrument will be 10$\times$ faster than SCUBA-2 to achieve a given sensitivity in a map. In the best case, the mapping speed can be increased by a factor of 20. The much improved mapping speeds of the new camera will enable deep and extensive continuum surveys (see section \ref{surveys}) within reasonable telescope time.

\subsubsection{The capability of the new camera: Recovering more extended emission}

\begin{figure*}
  \centering
  \includegraphics[width=0.95\textwidth]{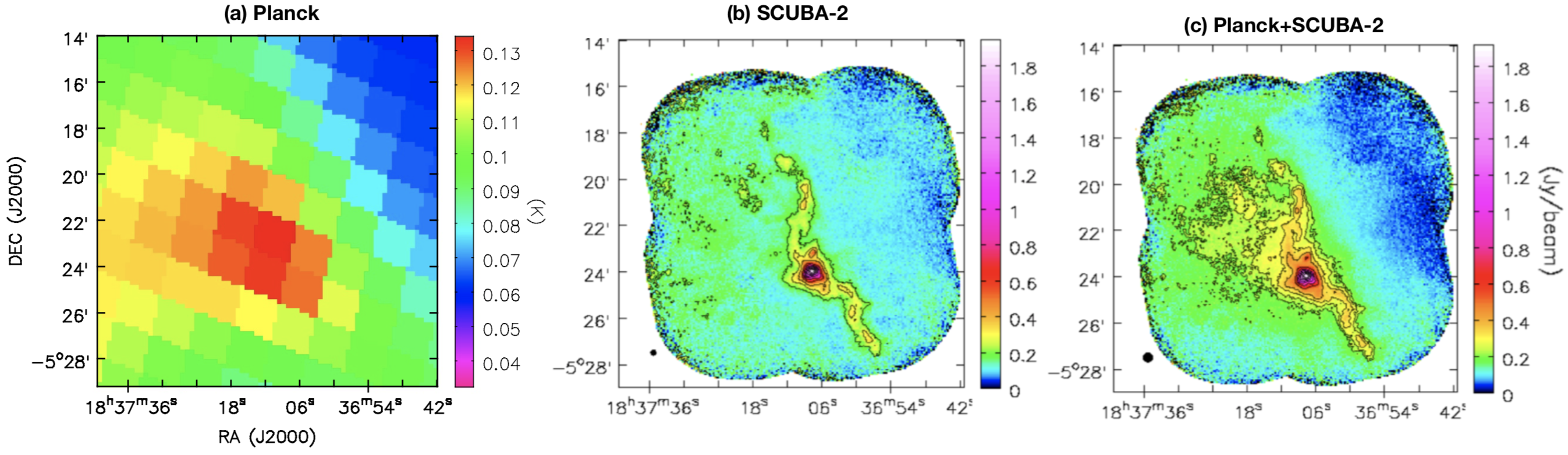}
  \caption{Images for PGCC G26.53+0.17 \citep{Liu2018a} (a) Planck 353 GHz data only (b) SCUBA-2 850 $\mu m$ data only (c) Combined SCUBA-2 and Planck data}
  \label{fig:G26}
\end{figure*}

The new camera will have the ability to recover more extended emission than the current camera SCUBA-2 benefiting from its improved pixel sensitivity, larger FOV and much faster scan speed. These data can be combined with Planck observations at 353 GHz in the Fourier domain to recover emission from all scales of molecular clouds \citep{Lin2017}. Panel (c) in Fig. \ref{fig:G26} presents one example of combining Planck data with SCUBA-2 data obtained from the SCOPE survey \citep{Liu2018a}. The combined data resolves the Planck clump (panel a) and recovers more extended emission than the SCUBA-2 data alone (panel b). With the ability to recover more extended emission with the new camera, the process to recover the extended emission will be more accurate in the planned surveys mentioned in section \ref{surveys}. 

\begin{figure*}[tbh!]
  \centering
  \includegraphics[width=0.8\textwidth]{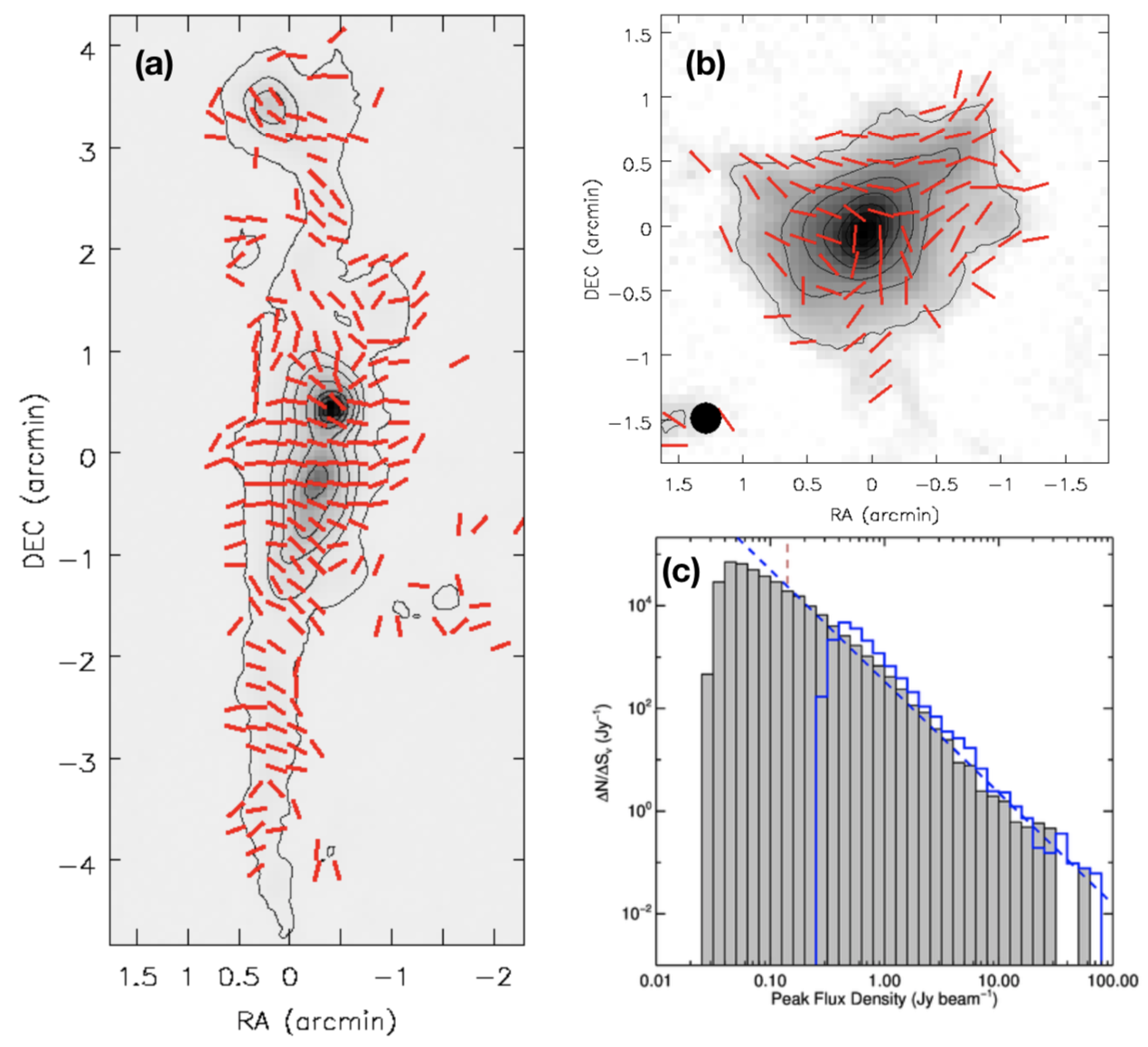}
  \caption{\textit{\textbf{(a)} JCMT POL-2 observations of a bright filamentary cloud (Soam, A. et al., 2019, submitted). The line segments show the inferred magnetic field morphology with a $S/N>2$. The contours are [0.1, 0.5, 1, 2, 4, 8] Jy~beam$^{-1}$. \textbf{(b)} JCMT POL-2 observations of a bright high-mass star-forming clump \citep{Liu2018c}. The segments and contours are the same as in panel (a). \textbf{(c)} Peak flux density distributions for the JPS (grey-filled histogram) compared to the ATLASGAL distribution (blue histogram)\citep{Eden2017}. The dashed line indicates the least-squares fit to the JPS distribution.} }
  \label{fig:Bfields}
\end{figure*}

\subsubsection{The capability of the new camera: Intrinsic polarization mapping capability}

Each pixel of the new camera is comprised of two detectors, that measure orthogonal linear polarization. By careful choice of the orientation of pixels across the focal plane, it will be possible to determine all Stokes parameters from a single scan observation, without the need for a rotating half-wave plane. This would enable us to conduct dust polarization observations simultaneously in the proposed surveys. In contrast, the aforementioned previous surveys except the Planck survey did not have polarization information. Limited by its poor resolution, Planck did not provide resolved polarization information within clouds.

The proposed surveys in Section \ref{surveys} have desired sensitivities of 2-10 mJy~beam$^{-1}$ in Stokes I to detect faint dust emission. This sensitivity will also capture dust polarization toward the bright filaments, clumps, and cores at the same time. Panels (a) and (b) of \textit{Fig. \ref{fig:Bfields}} show the results of SCUBA-2/POL-2 observations of one bright filamentary cloud and one high-mass star-forming clump, respectively, whose mean surface brightness is $\sim$1 Jy~beam$^{-1}$. The on-source time of those observations is $\sim$2 hrs with a sensitivity in Stokes I of $\sim$5-10 mJy~beam$^{-1}$. The new camera will achieve comparable sensitivity and map coverage in about 10 minutes of on-source integration. As shown in Panel (c) of \textit{Fig. \ref{fig:Bfields}}, hundreds of clumps discovered in the previous JPS and ATLASGAL surveys show peak flux densities larger than 1 Jy~beam$^{-1}$. Those clumps can be easily detected in polarization observations assuming polarization fractions of 1\%. It is also possible to detect polarization in less bright sources by smoothing the Q and U maps.

\subsection{New Galactic continuum surveys}
\label{surveys}

Benefiting from its improved pixel sensitivity, larger FOV, much faster scan speed and intrinsic polarization mapping capability, extensive Galactic continuum surveys with the new camera will not only detect the entire spectrum of molecular clouds throughout the Galaxy but also provide their resolved polarization information. To this end, we propose the following Galactic continuum surveys with the new camera in next decade.

\subsubsection{An unbiased continuum survey of the local volume}
\label{PCCs survey}

The sky visible to the JCMT is a declination band in the range $-30^{\circ}\leq\delta\leq+70^{\circ}$, encompassing some 18,000 square degrees. A blind deep survey, however, is still impossible with the new camera. 

Fortunately, the Planck satellite has done an all-sky multi-band continuum survey and discovered thousands of compact sources \citep{Planck2016b}. There are 9677 Planck compact objects detected at 353 GHz that are visible to the JCMT. This sample contains both cold and warm/hot Galactic clumps. These objects are not included in the proposed Galactic Plane survey in Section \ref{GPsurvey} and represent an unbiased sample of (sub)mm continuum sources distributed in the local volume ($<$2 kpc) \citep{Planck2016,Planck2016b}. The Planck catalogue serves as a nice guide which makes the survey of dense cores in all the visible local volume ($<$1-2 kpc) possible. 

The Planck observations of these objects, however, are at very low resolution ($\sim$5 arcmin) and do not resolve the structure of these sources.  With the new camera, all 9677 of these objects can be followed up at 14-arcsec resolution (over a factor of 20 improvement) in CV Daisy mapping mode\footnote{https://www.eaobservatory.org/jcmt/instrumentation/continuum/scuba-2/observing-modes/\#CV\_Daisy} with a total time of $\sim$320 hrs and a typical rms level of 5 mJy~beam$^{-1}$. In addition, the larger FoV of the new camera will make the CV Daisy scan naturally larger on the sky and it will extend the central area from 3 arcmin (SCUBA-2) to at least 6 arcmin, where the map will have uniform coverage.

Pilot HI observations of PGCCs indicate that half of them show narrow self-absorption features in HI spectra (see the right panel of \textit{Figure} \ref{fig:HI}; Tang et al. 2019, submitted). These objects are particularly interesting for studying the formation and evolution of molecular clouds \citep{Zuo2018}. We plan to conduct a deep (2 mJy~beam$^{-1}$) continuum survey of 1000 PGCCs that have been observed with $^{12}$CO/$^{13}$CO J=1-0 lines in the TOP-SCOPE survey \citep{Liu2018a} and will be mapped in HI with the Five-hundred-meter Aperture Spherical radio Telescope (FAST). At this high sensitivity, we can detect dust emission in the warm ($\sim$40 K) and low density (A$_V\sim$1; or 850 $\mu$m flux of $\sim$10 mJy) cloud peripheries, where HI is converted to H$_2$.  With the new camera, we need 0.2 hrs per source to achieve the desired sensitivity. The total time to complete this survey is about 200 hrs in Band 3 weather or 130 hrs in Band 2 weather. 

Besides the surveys of Planck clumps, a re-visit to the Gould Belt cloud with the new camera is also very meaningful considering no polarization information provided by either Herschel GBS or JCMT GBS. Herschel GBS\footnote{http://www.herschel.fr/cea/gouldbelt/en/} has a much larger sky coverage (160 deg$^2$) than the previous JCMT GBS (50 deg$^{2}$). The clouds in Herschel GBS span a range of physical conditions, from active, cluster-forming complexes to quiescent regions with lower star formation activity. A future generation JCMT GBS could cover all clouds in the Herschel fields with a high and uniform sensitivity ($\sim$5 mJy~beam$^{-1}$). With the new camera, it will take about 400 hrs in Band 3 weather or 250 hrs in Band 2 weather to complete a survey of the whole 160 deg$^2$ Herschel fields with the Pong 7200 mapping mode. The large spatial dynamic range of the proposed JCMT GBS in conjunction with Herschel GBS will probe the link between large-scale diffuse structures and dense structures (e.g., filaments or self-gravitating cores).

\subsubsection{Galactic Plane survey}
\label{GPsurvey}

Observations of external galactic systems, especially grand-design structures, have shown that the majority of star formation is occurring within the spiral arms of galaxies, galaxies like the Milky Way \citep[e.g.,][]{Urquhart2014}. The spiral arm structure within the Milky Way is confined to what is classically thought of as the Galactic Plane, although there is evidence of spiral arms leaving the mid-plane at extended Galactocentric radii \citep{DameThaddeus2011}. To understand star formation within the Milky Way and the extended Universe, we need to first understand how Galactic environments, ranging from large-scale features such as the spiral arms and the Galactic Bar to local triggering agents such as H{\sc ii} regions or star formation feedback, impact the star-formation process. 

As shown in the right panel of \textit{Fig. \ref{fig:SCOPE}}, a large number of Planck compact objects are distributed beyond the Galactic Plane. These sources indicate that star formation activities may occur in mid-latitude or high-latitude clouds. There are also many clumps at latitudes $1^{\circ}<|b|<2^{\circ}$ that were not covered by previous Galactic Plane surveys. Filling in these gaps was the original intent of the ``JPS'' and ``SASSy'' surveys before being descoped. A new high resolution ($\sim$10 arcsec) and sensitivity ($<$10 mJy~beam$^{-1}$) sub-millimeter Galactic Plane survey is needed to resolve the confusion in Hershel/Hi-GAL/Planck data and to detect objects in regions beyond the previous Galactic Plane surveys.

The new high-sensitivity MKID camera makes a ``full'' and wider Galactic Plane survey possible. One plausible Galactic Plane survey strategy is to observe the inner Plane ($-5^{\circ}<l<60^{\circ}$) with a wide latitude range ($|b|<5^{\circ}$; as wide as the Milky Way Imaging Scroll Painting (MWISP) project\footnote{http://english.dlh.pmo.cas.cn/ic/in/}) and to observe the remaining Plane ($60^{\circ}<l<240^{\circ}$) visible to the JCMT with a narrower latitude range ($|b|<2^{\circ}$). The sky coverage of this survey is outlined by the green box in the left panel of \textit{Fig. \ref{fig:SCOPE}}. The desired rms level is $\sim$10 mJy~beam$^{-1}$ at 850 $\mu$m or a 3$\sigma$ mass sensitivity of $\sim$0.1 M$_{\odot}$ for a point source out to a distance of 1 kpc assuming 20 K dust with a $\beta$ of 2. With the new camera, it takes  $\sim$900 hrs in Band 3 weather or $\sim$600 hrs in Band 2 weather to complete such a sensitive Galactic Plane survey with the Pong 7200 mapping mode\footnote{https://www.eaobservatory.org/jcmt/instrumentation/continuum/scuba-2/observing-modes/\#2-degree\_Pong\_map}.

\subsubsection{Science goals of the new Galactic continuum surveys}

The above surveys will provide a complete and unbiased census of molecular clouds, filaments, and dense clumps/cores in the local volume and in the Galactic Plane. JCMT has better resolution than the Herschel/SPIRE instrument and the new camera has ability to recover more extended emission than SCUBA-2. This will help resolve the widths of Herschel filaments in nearby ($<$1 kpc) clouds. Adding 850 $\mu$m data in SED fits will better constrain temperature and density profiles for dense filaments and also dense cores. From these surveys, we will thoroughly investigate star formation across the Galaxy, adressing the following goals:

\begin{itemize}

\item To build a complete database of molecular clouds in the Milky Way. To analyze molecular cloud distributions and properties across a range of Galactic environments (e.g. spiral arms, interarms, median/high latitude, expanding H{\sc ii} regions, supernova remnants). To compare Galactic molecular clouds with molecular clouds in nearby galaxies that are resolved in interferometric observations (e.g. ALMA). By combining JCMT data with Planck data, we will be able to detect faint and extended dust emission at the warm and low density cloud peripheries, where HI is converted to H$_2$. \textbf{Through a combined analysis of HI narrow self-absorption, CO emission, and dust emission, we will study the formation and evolution of molecular clouds. }

\item The dense gas fraction in molecular clouds can be estimated from the 850 $\mu$m flux ratio between the JCMT data (that mainly traces dense gas) and the JCMT+Planck combined data (that recovers extended gas emission). Then we will evaluate whether extended or diffuse emission (non star forming gas) dominates the 850 $\mu m$ flux in molecular clouds, and how does the dense gas fraction vary as a function of latitudes and longitudes. In particular, we will check whether or not there is a universal relation between star/core formation rate and surface mass density above a density threshold (or dense gas star formation law). 

\item In contrast to Herschel, JCMT observations have higher resolution and less confusion, and thus can more easily resolve filament widths and profile shapes. For example, assuming a typical filament width of 0.1 pc, the new camera will spatially resolve filaments beyond 1 kpc, whereas Herschel observations are generally restricted to 500 pc. In conjunction with molecular line Galactic Plane surveys (e.g., CHIMPS2\footnote{https://www.eaobservatory.org/jcmt/science/large-programs/chimps2/}, FUGIN\footnote{https://nro-fugin.github.io}, MWISP), we will conduct a complete census of velocity-coherent dense filaments in the Galactic Plane. We will detect filaments across the Galaxy and determine how universal they are in the ISM and how they are distributed relative to large-scale Galactic features (e.g., spiral arms and shells).

\item For resolved filaments in nearby ($<$1 kpc) clouds, we will investigate the distributions of their masses, line masses, lengths and widths. We will study how these filament parameters change in different clouds.  \textbf{Particularly, we will determine whether or not there is a characteristic width ($\sim$0.1 pc) of filamentary substructures in the ISM. }

\item To investigate the role of magnetic fields in filament formation by studying the relative orientations between filaments and large-scale magnetic fields, and how the field structures change within filaments. To check whether magnetic fields are important in supporting filaments or clumps/cores against gravitational collapse. To study the fragmentation process of filaments and to determine which physical processes (magnetic fields, turbulence or gravity) dominate their fragmentation.

\item To construct a complete sample of dense cores (with sizes of $\sim$0.1 pc) in clouds in the local volume ($<$2 kpc). We will then thoroughly study how core/clump mass functions and the core/clump formation efficiency change in different kinds of cloud. In particular, the new camera will be able to detect fainter cores with smoother structures in nearby clouds (e.g., Taurus) that were missed in previous SCUBA-2 surveys (e.g. SCOPE survey \citep{Liu2018a}). \textbf{The new surveys will provide a complete sample of dense cores for studying core evolution from very flattened starless phase to the centrally peaked protostellar phase.}

\item To construct a complete catalogue of high-mass clumps ($M>100~M_{\odot}$, size$\sim$1 pc) throughout the whole Galactic Plane. To derive the clump mass function, clump formation efficiency and the fraction of clustered vs isolated high-mass core (or star) formation across the Galaxy. To evaluate the gravitational stability of a large ($\times$1000) sample of high-mass clumps through virial analysis by considering all supports from magnetic fields, turbulence and thermal pressure. 

\item To detect (sub)mm transient objects through multi-epoch observations \citep{Park2019}. 

\item In particular, these surveys will provide a unique complete catalogue of dense cores/clumps for follow-up high-resolution molecular line/dust continuum studies with other mm/submm telescopes and interferometers (e.g., ALMA).

\end{itemize}

\end{document}